# Analysis of the magnetic and magnetocaloric properties of ALaFeMnO$_6$ (A=Sr, Ba and Ca) double perovskites

N. Brahiti[1], M. Abbasi Eskandari[1], M. Balli[1,2], C. Gauvin-Ndiaye[1], R. Nourafkan[1], A.-M.S. Tremblay[1], P. Fournier[1,a]

[1]Institut quantique, Regroupement québécois sur les matériaux de pointe et Département de physique, Université de Sherbrooke, Sherbrooke, J1K 2R1, Québec, Canada

[2] LERMA, ECINE, International University of Rabat, Parc Technopolis, Rocade de Rabat-Salé, 11100, Morocco

[a]Corresponding author and electronic mail: patrick.fournier@usherbrooke.ca

**ABSTRACT**

In previous studies, we have reported that double perovskite La$_2$NiMnO$_6$ presents a non-negligible potential for room temperature magnetocaloric tasks. With the aim of improving even further the cooling performances and the working temperature range of double perovskites, we report the magnetic and magnetocaloric properties of La$_2$MnFeO$_6$ and ALaMnFeO$_6$ (A=Sr, Ba, Ca) compounds. X-ray diffraction (XRD) and Rietveld refinement show that La$_2$MnFeO$_6$ (LMFO) and CaLaMnFeO$_6$ (CLMFO) samples crystallize in an orthorhombic structure with the *Pnma* space group. However, a rhombohedral structure with the $R\bar{3}C$ space group is obtained for BaLaMnFeO$_6$ (BLMFO) and SrLaMnFeO$_6$ (SLMFO) samples. Substituting La by Ba or Sr in LMFO leads to a clear increase of the Curie temperature (T$_c$) compared to LMFO from 150 K for BLMFO up to 350 K for SLMFO. Moreover, CLMFO shows the smallest T$_c$ down to 70 K. Ferromagnetic-like behavior is observed for SLMFO and BLMFO while CLMFO's magnetism resembles that of LMFO. A clear connection between the structural parameters and the magnetic properties of these doped LMFO samples is unveiled as the highest T$_c$ and the largest magnetization are observed for SLMFO which shows also bond angles closest to 180º and the smallest bond lengths, thus optimizing the superexchange interaction. The partial substitution of Sr for La leads in fact to a significant magnetocaloric effect over a wide operating temperature range extending beyond 300 K. For some optimal growth conditions, its entropy change varies slowly over an unusually large temperature range, which is of clear interest from a practical point of view.

**INTRODUCTION**

Based on the magnetocaloric effect (MCE), magnetic refrigeration is an environment-safe energy-efficient refrigeration alternative to conventional systems based on the fluid compression-expansion process [1-5]. The MCE phenomenon has been used for many years to reach very low temperatures in a process also known as adiabatic demagnetization using paramagnetic salts [1-5]. Starting in 1976, magnetic refrigeration near room temperature has stirred a lot of interest when Brown unveiled an innovative magnetocaloric device working with gadolinium metal as the active refrigerant [6].

With the goal of improving the general performances of actual prototypes, search for novel magnetocaloric materials led to the discovery of a giant MCE near room temperature by Pecharsky and Gschneidner in $Gd_5Si_2Ge_2$ [7]. Following this breakthrough, a large variety of advanced magnetocaloric materials was proposed for room temperature magnetic cooling [1, 8-18]. In recent years, this search has partially focused on the exploration of promising new ferromagnetic oxides for which magnetic coupling is driven either by the superexchange or the double exchange mechanisms [1, 4, 16, 17].

From this group, double perovskites of general formula $R_2BB'O_6$ (R = rare earth; B, B' = pair of 3d metals) have recently attracted much attention. In addition to their desirable magnetocaloric properties, these oxides possess some key assets when compared to intermetallics, such as high electrical resistivity, chemical stability and a low production cost [1, 4]. For instance, it was shown previously that the cation-ordered phase of the ferromagnetic insulator $La_2NiMnO_6$ (LNMO) presents a second-order magnetic phase transition at $T_C \sim 280$ K leading to significant MCE levels at room temperature [15, 18]. Moreover, it was also shown that the Curie temperature of LNMO strongly depends on the level of Ni and Mn cation ordering in its crystal structure since two regions of the material with different crystal symmetries and level of cation ordering could be identified using, for example, transmission electron microscopy [19]. The cation-disordered phase of LNMO presents in fact a low-temperature transition. The presence of nanodomains with different transition temperatures in specific samples can in fact be used to tailor their magnetic response to suit the requirements for efficient cooling by controlling the relative amount of the two phases [18]. Although LNMO qualifies most probably as a prime candidate to be tested as an active material in existing prototypes, one would like to further improve the properties by substituting either the rare earth La, or one of the *3d* metal atoms, Ni or Mn in order to achieve transition temperatures above room temperature and tunable MCE.

For this purpose, two kinds of double perovskites were synthesized by the conventional solid-state reaction technique: 1) a total substitution of Ni in LNMO by Fe to form $La_2MnFeO_6$ (LMFO), a double perovskite which we expected initially to exhibit strong ferromagnetic exchange interactions; 2) a partial substitution of La in LMFO by divalent cations to form the compound $ALaMnFeO_6$ (A=Sr, Ba and Ca). Below, we show that $La_2MnFeO_6$ is not ferromagnetic carrying an unexpectedly low magnetization, in agreement with previous studies. In fact, this behavior can be explained using density functional theory (DFT) by an antiferromagnetic superexchange coupling between Fe and Mn conditioned by strong electron-electron interactions and Jahn-Teller distortions [20]. Following a prediction from the same theoretical study in Ref. 20 and a follow-up study in Ref. 21, the partial substitution of La by divalent alkaline earth metal ions leads to an enhancement of the magnetization and an increase in the transition temperature above room temperature suggesting a path to reach larger magnetocaloric effect than $La_2NiMnO_6$ at room temperature in doped double perovskites.

**METHODS**

Polycrystalline samples of $La_2MnFeO_6$ and $ALaMnFeO_6$ (ALMFO) compounds were synthesized by the standard solid-state reaction method. High-purity oxides or carbonates $La_2O_3$, $Fe_2O_3$, $MnO_2$, $SrCO_3$, $BaCO_3$ and $CaCO_3$ were used as starting materials and mixed in proper stoichiometric proportions. The mixtures were ground in a mortar until homogeneous powders were obtained and then calcined in air for 24 h at different temperatures. The LMFO powder was heated from 900°C to 1100°C by steps of 50°C with intermediate regrinding while ALMFO powders were instead sintered up to 1050°C. All of them were then finally pressed into pellets of 12 mm diameter and 2 mm thickness and sintered for 24h at 1070°C for ALMFO and 1150°C for LMFO.

The crystal structure of these compounds was then confirmed using powder x-ray diffraction (XRD) measurements with a Bruker-AXS D8-Discover diffractometer in the θ-2θ configuration with a Cu Kα source over the 2θ range of 10 to 80°. The structural parameters were obtained by fitting the experimental XRD data using the Rietveld structural refinement FULLPROF software applying the pseudo–Voigt peak shape function and a linear interpolation for background description. The refinements were performed until convergence as shown by the goodness of fit ($\chi^2$). The magnetization measurements were performed with a Superconducting quantum interference device (SQUID) magnetometer or the ACMS option of a Physical properties measurements system both from Quantum Design. Density functional theory (DFT+U) calculations were performed within the full potential all-electron basis set as implemented in the WIEN2k code. The calculations were used to gain insight on the crystal-field splitting of the transition metals d-orbitals of the various compounds. Total-energy calculations allowed to determine the magnetic ground state of each compound and Monte Carlo simulations of the effective Ising model were used to estimate their transition temperature. Details of these calculations have been presented in separate papers [20, 21]. This approach helped us to understand the origin of magnetic properties that cannot be understood with a naive toy model for superexchange in these materials, especially for LMFO. It was also used to make predictions on how to alter their magnetic properties and improve their magnetocaloric performances.

**RESULTS AND DISCUSSION**

The room temperature powder x-ray diffraction (XRD) patterns of $La_2MnFeO_6$ (LMFO), $SrLaMnFeO_6$ (SLMFO), $BaLaMnFeO_6$ (BLMFO) and $CaLaMnFeO_6$ (CLMFO) compounds are presented in Figure 1. The observed reflections in the XRD patterns correspond to single phase double perovskite oxides for all the samples. On the other hand, the broad reflections for the doped LMFO indicate somewhat smaller size crystallites than for the undoped compound. An example of a Rietveld refinement of the XRD data for SLMFO is shown in Figure 1(b) while the remaining fits are presented in the supplementary material. The data and the refinements reveal that LMFO and CLMFO adopt an orthorhombic (*Pnma*) structure while a rhombohedral structure (*R-3c*) is obtained for the SLMFO and BLMFO compounds in partial accordance with previous reports [22, 23]. Shaheen *et al.* reported an orthorhombic perovskite structure, but with space group *Pbnm* for BLMFO and CLMFO, while a rhombohedral structure (space group *R-3c*) for SLMFO was observed [22]. More recently, Kumar *et al.* reported the structural analysis and optical properties of BLMFO and SLMFO samples prepared by the auto-combustion method [23]. They found that BLMFO may be indexed to a cubic crystal structure with space group *Pm-3m* while SLMFO crystallizes into the rhombohedral symmetry within the space group *R-3c* in agreement with Ref. [22].

Table 1 presents a summary of the crystal structure parameters and the unit cell volume extracted from the refinements and their correlation to the ionic radii. In this table, the average length of the (B,B')-O bonds, the average bond angle of B-O-B' (B,B' = Mn or Fe) and the goodness of fit $\chi^2$ are also reported. The unit cell volume is primarily determined by the ionic radii of A, R, B and B' cations in the $ARBB'O_6$ double perovskite. Moreover, the changing bond angle in the respective unit cell are driven by lattice distortions arising from the mismatch of the (A,R)-O and (B,B')-O sublattices bond lengths. This mismatch can be quantified using the tolerance factor ($t_f$) defined as $t_f = \frac{r_{A,R}+r_O}{\sqrt{2}(r_{B,B'}+r_O)}$ where $r_{A,R}$ is the average ionic radius of cations on the R(A) site, $r_{B,B'}$ is the average ionic radius of cations on the B and B' sites and $r_O$ the ionic radius of $O^{2-}$, respectively. In our case, we notice that the lattice parameters, *a*, *b* and *c* and the unit cell volume are correlated to the size of the ionic radius of the dopant (Ca, Sr, Ba) for the doped samples. This is especially underlined by the volume per formula unit scaling clearly with the respective radii. However, the unit cell volume of CLMFO compared to LMFO can only be explained by the smaller volume of the $Mn^{4+}O_6$ octahedra compared to that of $Mn^{3+}O_6$ in LMFO. The calculated values of the tolerance factor from the above relation using the expected (average) ionic radii, $t_{fcal}$, and the observed tolerance factor determined from Rietveld refinements $t_{fobs}$ are listed in Table 1. According to the calculated values of $t_{fcal}$, orthorhombic and rhombohedral distortions are expected for LMFO and doped LMFO, respectively. The crystal cell parameters calculated with the Rietveld refinements fit perfectly with our experimental data for LMFO, SLMFO and BLMFO in the expected crystalline structures, while the best fit of CLMFO which gives the smallest value of goodness of fit ($\chi^2$) is obtained with an orthorhombic structure. We can notice in Table 1 that the largest discrepancies for the tolerance factors occur for LMFO and CLMFO. Moreover, we notice that the B-O-B' average bond angle is significantly smaller in CLMFO when compared to LMFO, SLMFO and BLMFO.

**Table 1**: Crystal structure parameters extracted from the Rietveld refinements.

| Compounds | LMFO | CLMFO | SLMFO | BLMFO |
|---|---|---|---|---|
| **Ionic radii (Å)** | $La^{3+}$ : 1.36 | $Ca^{2+}$ : 1.34 | $Sr^{2+}$ : 1.44 | $Ba^{2+}$ : 1.61 |
| **Space group** | Pnma | Pnma | R-3c | R-3c |
| **a (Å)** | 5.5423 | 5.4773 | 5.4732 | 5.5478 |
| **b (Å)** | 7.8188 | 7.6694 | 5.4732 | 5.5478 |
| **c (Å)** | 5.5080 | 5.4224 | 13.4919 | 13.5265 |
| **V (Å$^3$)** | 238.68 | 227.77 | 350.01 | 360.54 |
| **V per f.u. (Å$^3$)** | 119.34 | 113.885 | 116.67 | 120.18 |
| **$d_{B,B'-O}$ (Å)** | 1.975 | 2.110 | 1.93 | 1.971 |
| **$\Theta_{B-O-B'}$ (°)** | 160.12 | 129.78 | 178.43 | 168.91 |
| **$t_{f\,cal}$** | 0.969 | 0.978 | 0.996 | 1.026 |
| **$t_{f\,obs}$** | 0.949 | 0.943 | 0.982 | 1.0005 |
| **$\chi^2$ (%)** | 1.604 | 1.134 | 1.381 | 1.830 |

Figure 2 shows the zero-field cooled (ZFC) magnetization as a function of temperature measured under an applied magnetic field of 0.2 T for undoped and doped LMFO. These curves are used mainly to pinpoint the transition temperature and guide our choices of temperatures for isotherm measurements. The measured magnetization curve of LMFO is like that reported by Barrozo *et al.* [24]. A broad transition centered around 80 K is observed for LMFO, presumably from a ferromagnetic (FM) to a paramagnetic (PM) phase. Figure 3(a) presents the full hysteresis curves of LMFO, CLMFO, BLMFO and SLMFO samples at 5 K. The observed magnetization value of roughly 1.8 $\mu_B$/f.u. at 7T, 5K for LMFO is smaller than the expected values of 7 $\mu_B$/f.u. for $Mn^{+3}/Fe^{+3}$ or 9$\mu_B$/f.u. for $Mn^{+4}/Fe^{+2}$ configurations if one assumes a ferromagnetic coupling between the Mn and Fe moments. The low magnetization value of LMFO could be attributed to a possible B/B' cationic disorder leading to the formation of a random distribution of Mn-O-Fe, Mn-O-Mn and Fe-O-Fe bonds or even regions with only Mn-O-Mn or Fe-O-Fe bonds [24]. In this case, the $Fe^{+3}$-O-$Mn^{+3}$ or $Fe^{+2}$-O-$Mn^{+4}$ bonds, presumably both ferromagnetic, are diluted in a matrix made of $Mn^{+3}$-O-$Mn^{+3}$ or $Fe^{+3}$-O-$Fe^{+3}$ antiferromagnetic bonds. The result is a low temperature transition with small magnetization. Another possible scenario is that the cation-ordered phase is antiferromagnetic and, even though it is dominant, it leads to a low magnetization.

In fact, recent DFT calculations assuming B/B' cationic order (with only Fe-O-Mn bonds) indicate that the Mn and Fe moments are coupled antiferromagnetically and that the lowest energy bond corresponds to oxidation states given by $Fe^{+3}$-O-$Mn^{+3}$ [20]. This study revealed the important role played by the strong electronic repulsion for electrons in the Fe-*3d* orbitals determining the electronic ground state of LMFO at the origin of the antiferromagnetic exchange. Indeed, these strong electron-electron interactions prevent the double occupancy of the Fe-*3d* orbitals and lead to a $Fe^{3+}$ state with half-filled *3d* orbitals and a $Mn^{3+}$ state. The subsequent Jahn-Teller distortion arising from the $Mn^{3+}$ ions lifts the degeneracy of the Mn-$e_g$ states. The interplay between the Jahn-Teller distortion and the strong Fe-*3d* correlations then leads to an antiferromagnetic oxygen-mediated superexchange interaction between neighboring $Mn^{3+}$ and $Fe^{3+}$ moments. This interaction is illustrated schematically in Figure 4(a), where we assumed that the $Mn^{3+}$ ion has a $3d_\sigma^4 d_{\bar\sigma}^0$ configuration. In that case, the Jahn-Teller distortion lifts the degeneracy of the Mn-$e_g$ orbitals and favors the hopping of an O-p spin-down electron to a Mn-$e_g$ spin-down orbital. The remaining O-p spin-up electron can only contribute to the superexchange interaction if the high-spin $Fe^{3+}$ ion has a $3d_\sigma^0 d_{\bar\sigma}^5$ configuration. Hence, the antiparallel alignment of neighboring $Mn^{3+}$ and $Fe^{3+}$ moments leads to a higher kinetic energy advantage than their parallel alignment, which explains the AFM ground state of LMFO. As a result, one expects that the maximum magnetization of ordered LMFO could theoretically reach 1$\mu_B$/f.u. in fair agreement with our measured high-field data for LMFO in Figure 3(a).

The same theoretical study based on DFT calculations made also the prediction that partial La substitution by Ca, Sr and Ba should lead to ferromagnetism for the (La/A)-site ordered CLMFO, SLMFO and BLMFO [21]. In fact, the divalent substitution of half of the La atoms should lead to hole doping into Mn-O-Fe covalent bonds, i.e. to $Mn^{4+}$ and $Fe^{3+}$ states without Jahn-Teller distortion or double occupancy of Fe-$3d$ orbitals. The ferromagnetic ground state should show a total moment of 8$\mu_B$/f.u. arising from the $Mn^{4+}$ and $Fe^{3+}$ ions [19]. The ferromagnetic interaction between neighboring $Mn^{4+}$ and $Fe^{3+}$ moments is then easily understood from a simple superexchange model. This is once again illustrated in Figure 4(b), in which we assumed that the $Mn^{4+}$ ion has a $3d_\sigma^3 d_{\bar\sigma}^0$ configuration. In this case, an O-p spin-up electron can hop on the Mn-$e_g$ orbital due to Hund's coupling. For the remaining O-p spin-down electron to contribute to the superexchange interaction, high-spin $Fe^{3+}$ must have a $3d_\sigma^5 d_{\bar\sigma}^0$ configuration. Thus, the kinetic energy advantage is larger if the neighboring $Mn^{4+}$ and $Fe^{3+}$ moments have a parallel alignment, resulting in a FM ground state [21].

Consequently, substituting La by Ba or Sr in LMFO leads to a clear increase of the Curie temperature to 150 K and up to 350 K, respectively, as shown in Figure 2. However, Ca substitution leads to a behavior like LMFO. A partial substitution of the trivalent $La^{3+}$ by a divalent cation ($Ca^{2+}$, $Sr^{2+}$, $Ba^{2+}$) must change the oxidation state of either Mn or Fe. This change appears to have a very deep impact on the stable magnetic phase probably through a (partial) change in the nature of the exchange interaction. Nevertheless, since these divalent cations should provide the same number of holes to the structure, the observed differences between CLMFO, SLMFO and BLMFO signal the likely impact of lattice distortions driven by the changes in the tolerance factor quantifying the ratio of the (La/A)-O average bond length with respect to the (Mn,Fe)-O average bond length and the B–O–B' average bond angle as another way to control the strength of the superexchange interaction. From the Table 1, the highest bond angle is obtained for SLMFO leading to the strongest superexchange interaction, followed by BLMFO, LMFO then CLMFO. These results are in accordance with the obtained Curie temperatures and the magnetization values of the doped samples. Further studies and characterizations are needed to understand the contrasting behavior of Ca doping with respect to Sr and Ba doping, but one must notice that the extreme bond angle for CLMFO estimated from the Rietveld refinements (see Table 1) as well as its fairly elongated (B,B')-O average bond length should play a crucial role in determining the rather poor magnetic response (low $T_C$ and low magnetization).

Comparing the full hysteresis curves of LMFO, CLMFO, BLMFO and SLMFO at 5 K in Figure 3(a), we observe weak hysteresis with low coercive field ($H_C$) values of about 250 Oe for all the samples. It is worth mentioning that the magnetization does not reach saturation and increases continuously at high magnetic field for all the samples. In addition, a low magnetization value is observed for all the samples even at our highest magnetic field of 7T. At best, it reaches 3.5$\mu_B$/f.u. for SLMFO, 2.8$\mu_B$/f.u. for BLMFO, and 1 $\mu_B$/f.u. for CLMFO. These magnetization values are lower than the theoretical saturation magnetization value of 8 $\mu_B$/f.u. if we assume a ferromagnetic coupling between $Mn^{4+} - O - Fe^{3+}$ ions for ALMFO (A = Sr, Ba, Ca) samples. Their low magnetization can be attributed in part to B/B' cationic disorder. In such condition, the contribution from the ferromagnetic superexchange-driven $Mn^{4+} - O - Fe^{3+}$ bonds are interrupted by a large proportion of antiferromagnetic $Mn^{4+} - O - Mn^{4+}$ and $Fe^{3+} - O - Fe^{3+}$ bonds. According to Goodenough *et al.*, ordering of $Mn^{4+}$ and $Co^{2+}$ ions in a similar double perovskite, $La_2MnCoO_6$ (LCMO), leads to a ferromagnetic $e^2 - O - e^0$ interactions [30]. These authors note that the cationic disorder creates antiferromagnetic $Mn^{4+} - O - Mn^{4+}$ or $Co^{2+} - O - Co^{2+}$ bonds. These bonds generate antiphase boundaries separating ferromagnetic domains, thus affecting long-range ferromagnetic order [30]. The structural ordering of B/B' can be roughly estimated by the ratio of the experimental to the theoretical magnetic moments (δ), which should be unity for perfect cationic order [31]. In the present study, assuming a theoretical magnetic moment of 8 μ$_B$/f.u., we estimate δ = 0.43, 0.35 and 0.125 for SLMFO, BLMFO and CLMFO, respectively. As a comparison, Madhogaria *et al.* reported that their LCMO samples exhibit a ratio of δ = 0.95, which was attributed mainly to B/B' cationic disorder and possibly to oxygen vacancies [32]. In the case of LMFO and the doped compounds, our low values of δ indicate that the cationic disorder is more important than in LCMO. It also suggests that there may be ways to improve the samples by tuning the growth conditions in order to promote a larger proportion of cationic order. It is this ability to tune the magnetic properties (i.e. δ) that is so attractive in these double perovskites [18].

In the present study, the sintering temperatures of 1070°C for ALMFO and 1150°C for LMFO samples are smaller than those usually used for the synthesis of double perovskite oxides. Consequently, small size crystallites are present in our samples possibly explaining the superparamagnetic-like behavior. Such results were reported by Bhame *et al.* [25] for polycrystalline $LaMn_{0.5}Fe_{0.5}O_3$ synthesized by a low-temperature method. These authors explained these magnetic hysteresis curves by the presence of ferromagnetic and superparamagnetic phases in their samples whose contributions depend on the processing conditions. This behavior can be also attributed to an inhomogeneous mixture of FM and AFM phases in all the prepared samples. It emphasizes a very interesting characteristic of double perovskites: it is possible to tune their magnetic properties by varying the growth conditions like what was shown recently with LNMO thin films [18]. Figure 3(b) shows the magnetic field dependence of the magnetization at 300K for SLMFO material where a ferromagnetic-like trend persists up to room temperature and beyond. Its magnetization value under a magnetic field of 0.5 T is about 0.3$\mu_B$/f.u., a low value with respect to the magnitude usually encountered in some of the best room temperature magnetocaloric materials.

One characteristic signature of the magnetic properties of these compounds is their very broad transitions, as illustrated in Figure 2, which one can get by manipulating the doping and the growth conditions. If the magnetization is large enough, this may produce a large enough magnetocaloric effect in a wide temperature range (see below) suitable for a specific application. The focus of the present work concerns essentially a comparison of the MCE in LMFO and ALMFO (A= Sr, Ba and Ca) materials. The MCE is usually quantified as a change of temperature in adiabatic conditions ($\Delta T_{ad}$) or the change of magnetic entropy in an isothermal process ($\Delta S_M$). In this study, the isothermal process was used to evaluate the entropy change of the prepared samples subjected to an external magnetic field. According to Maxwell's relation, the magnetic entropy change $\Delta S_M$ for an applied magnetic field variation from 0 to H is given by [1, 26- 29]:

$$\Delta S_M(T, 0 \to H) = \mu_0 \int_0^H \left(\frac{\partial M}{\partial T}\right)_{H'} dH' \qquad (1)$$

Since the magnetization is usually measured at discrete magnetic fields and temperatures, the isothermal magnetic entropy changes in Eq. (1) is usually computed by transforming the integral into a discrete sum as follows:

$$\Delta S_M = \mu_0 \sum_i \frac{M_{i+1} - M_i}{T_{i+1} - T_i} \Delta H_i \qquad (2)$$

where $M_{i+1}$ and $M_i$ are the magnetization at temperatures $T_{i+1}$ and $T_i$, respectively, under a magnetic field H measured with an increment of $\Delta H_i$. Another way to approach this calculation is simply to realize that Eq. (1) computes the surface between two isotherms measured at $T_{i+1}$ and $T_i$ from 0 to H.

From isothermal M(H) curves like those shown in Figure 5 and Eq. (2), we have calculated the magnetic entropy variation (-$\Delta S_M$) displayed in Figure 6. We first notice that all the samples exhibit a maximum value around the magnetic transition temperature $T_C$. Since $T_C$ is ill-defined with the broad transitions in Figure 2, the curves cover an extended range of temperature. In addition, the magnitude of $-\Delta S_M$ increases while increasing the external magnetic field for all the samples. Figure 6 shows that the maximum magnetic entropy value for LMFO (Figure 6(a)) increases from 0.19 to 0.7 J/K kg for a magnetic field variation of 3 to 7 T, respectively. However, the maximum entropy changes for BLMFO in Figure 6(b) and SLMFO in Figure 6(c) are about 0.15 to 0.42 J/K kg and 0.15 to 0.48 J/K kg, respectively. The MCE in terms of the maximum magnetic entropy change for these materials is significantly lower than that presented by gadolinium (Gd) metal with 9.8 J/kg K [1] and oxides such as $La_2NiMnO_6$ (LNMO) and $La_2CoMnO_6$ (LCMO) single crystals with MCE value of 2 and 3.5 J/K kg, respectively, for magnetic field variation from 0 to 5T [1, 26, 27]. However, the very large temperature range of significant $-\Delta S_M$ with almost flat temperature-independent values could be appealing if it can be further increased.

A comparison between the estimated entropy changes as a function of temperature with magnetic field varying from 0 to 5 T for LMFO, BLMFO and SLMFO is shown in Figure 7. Interestingly, while LMFO and BLMFO show well-defined but broad maxima, SLMFO shows an entropy change slowly varying over a temperature range as large as 300 K. Of course, approaching a flat so-called *table-top* temperature dependence is of great practical interest as it is a key ingredient to achieve the best refrigeration performances based on different thermomagnetic cycles, such as the Ericsson and Brayton (active magnetic regenerative refrigeration) cycles [1,27].

The relative cooling power (RCP), which represents the cooling efficiency of a magnetic refrigerant and indicates the amount of heat transferred between the hot and cold sink in the refrigeration cycle, is defined as:

$$\text{RCP (S)} = -\Delta S_M \times \delta T_{FWHM}$$

where, $-\Delta S_M$ is the peak of isothermal entropy change and $\delta T_{FWHM}$ is full width at half maximum value of the temperature range. Only a very rough estimate of the RCP values with respect to the magnetic field for SLMFO can be obtained using the low temperature half of the broad peaks observed in Figure 6(c). The resulting RCP as a function of the applied magnetic field is shown in Figure 8. The RCP exhibits a linear behavior with the magnetic field. In addition, the value at 5 T is around 105 J/Kg. RCP values are slightly smaller than those usually observed in polycrystalline manganites. For example, 215 J/Kg was reported for 5T by Bingham *et al.* in charge-ordered $Pr_{0.5}Sr_{0.5}MnO_3$ manganite [33], while 165 J/Kg was estimated in $La_{0.8}Ca_{0.2}Mn_{0.99}Fe_{0.01}O_3$ manganite prepared by the sol-gel method [34] and 270 J/Kg in $La_{0.7}Ca_{0.26}Sr_{0.04}MnO_3$ synthetized by the Pechini sol-gel method [35].

The low MCE and cooling efficiency in these doped LMFO can be attributed mainly to the low magnetization values. SLMFO could become a good candidate for room temperature magnetic refrigeration if one can markedly enhance its magnetization towards its expected theoretical maximal value. Its magnetocaloric properties might be enhanced simply by increasing the grain size of their polycrystalline forms using optimum annealing conditions. On the other hand, the MCE shown by these compounds is usually much lower if compared to its theoretical limit assuming a fully polarized ferromagnetic phase. From this perspective, high magnetocaloric response may still be achieved in this family with further growth manipulations as a possible way to tune further the relative proportion and volume of cation-ordered domains. The impact of the growth conditions on the magnetic properties is currently under investigation. Finally, a deeper understanding of the electronic structure of double perovskites and its interplay with other crystallographic and magnetic order parameters should pave the way for the design of advanced double-perovskite compounds able to operate close to their MCE theoretical limit. These ceramic materials exhibit other advantages such as high chemical stability, high resistance to corrosion and oxidation, low cost and easy fabrication process and a high electrical resistance (absence of eddy currents). These physical properties are of great interest from a practical point of view.

**CONCLUSION**

In the present study, La$_2$FeMnO$_6$ and ALaFeMnO$_6$ (A= Sr, Ba, Ca) double perovskites have been successfully synthesized by a standard solid-state reaction method. The structural, magnetic and magnetocaloric properties of these samples have been studied and compared. The temperature dependence of the magnetization shows ferromagnetic-like behavior for all prepared materials, but with low magnetization values. A clear correlation between the structural and magnetic is observed as the largest ferromagnetic transition temperatures and magnetizations are observed for Sr and Ba doping corresponding to the largest bond angles and smallest Fe-O-Mn bond lengths. The magnetic isotherms likely reveal a superparamagnetic behavior for all the samples. The magnetocaloric effect in terms of entropy change was also estimated from measurements of magnetic isotherms for LMFO, BLMFO and SLMFO. Its maximum value at 5T was found to be about 0.5 J/kg K for LMFO and 0.3J/kg K for both BLMFO and SLMFO. For LMFO, an effective antiferromagnetic interaction between Mn and Fe predicted by DFT calculations is at play, leading to a low magnetization. Partial substitution of La by Ca, Sr and Ba in LMFO changes the oxidation state of Mn, leading to ferromagnetic Mn-O-Fe bonds. However, the low magnetization and MCE for these materials may be attributed to the anti-site disorder in the B and B' sites of the double perovskite structure. Due to a very broad magnetic transition extending over 300K, the magnetocaloric effect of SLMFO covers a wide operating temperature range of roughly 300 K with a change of entropy varying slowly with temperature. Further structural studies and additional exploration of the growth conditions of these doped LMFO samples are required to fully understand the origin of their magnetic properties and learn how to control their growth in order to suit the needs of specific magnetocaloric tasks.

15ACCEPTED MANUSCRIPT

Journal of Applied Physics

This is the author's peer reviewed, accepted manuscript. However, the online version of record will be different from this version once it has been copyedited and typeset.
PLEASE CITE THIS ARTICLE AS DOI: 10.1063/1.5144153

AIP Publishing

**SUPPLEMENTARY MATERIAL**

See supplementary material for the complete set of x-ray diffraction data analysis using Rietveld refinement and isotherms measured for all the samples not presented in the main manuscript.

**ACKNOWLEDGMENTS**

The authors thank M. Castonguay, S. Pelletier and B. Rivard for technical support. We thank M. Côté (U. de Montréal) for fruitful discussions. This work is supported by the Natural Sciences and Engineering Research Council of Canada (NSERC) under grants RGPIN-2014-04584, and RGPIN-2013-238476, the Canada First Research Excellence Fund, by FRQNT (Québec), by the Research Chair in the Theory of Quantum Materials and the Université de Sherbrooke. Numerical simulations were performed on computers provided by the Canada Foundation for Innovation, the Ministère de l'Éducation, des Loisirs et du Sport (Québec), Calcul Québec and Compute Canada.

**FIGURE CAPTIONS**

**Figure 1 – (a)** Powder XRD patterns for (from bottom to top) LMFO, BLMFO, CLMFO and SLMFO. **(b)** Example of a Rietveld refinement for SLMFO (black: experimental data; red: refinement fit; blue: difference; vertical lines: markers of predicted reflections).

**Figure 2 -** Magnetization as a function of temperature for LMFO and substituted LMFO samples under an applied magnetic field of 0.2 T.

**Figure 3 -** Hysteresis loops for (a) CLMFO, SLMFO, BLMFO and LMFO at 5 K, and (b) SLMFO at 300K.

**Figure 4 -** Schematic representation of the oxygen mediated superexchange mechanisms at play in (a) LMFO and (b) ALMFO (A=Ba, Sr or Ca) as deduced from DFT calculations [20, 21].

**Figure 5 –** Selected isothermal magnetization curves of SLMFO sample from 50 K to 350 K (shown at intervals of 10K for clarity) used to evaluate the isothermal entropy change. Isotherms were measured at a 5K interval.

**Figure 6 -** Temperature dependence of the magnetic entropy changes of (a) LMFO, (b) BLMFO and (c) SLMFO for different magnetic field variations.

**Figure7** - Temperature dependence of the isothermal magnetic entropy changes of LMFO, BLMFO and SLMFO for a field sweep from zero to 5 T.

**Figure 8 -** Variation of the relative cooling power as a function of the applied magnetic field for SLMFO compound.

**REFERENCES**

[1] M. Balli, S. Jandl, P. Fournier, A. Kedous-Lebouc, Appl. Phys. Rev. **4**, 021305 (2017).

[2] K. A. Gschneidner, Jr., V. K. Pecharsky, and A. O. Tsokol, Rep. Prog.Phys. **68**, 1479 (2005).

[3] X. Moya, S. Kar-Narayan, and N. D. Mathur, Nat. Mater. **13**, 439 (2014).

[4] M. H. Phan and S. C. Yu, J. Magn. Magn. Mater. **308**, 325 (2007).

[5] K. G. Sandeman, Scr. Mater. **67**, 566 (2012).

[6] G. V. Brown, J. Appl. Phys. **47**, 3673 (1976)

[7] V. K. Pecharsky and K. A. Gschneidner, Jr., Phys. Rev. Lett. **78**, 4494 (1997)

[8] Y. F. Chen, F. Wang, B. G. Shen, F. X. Hu, Z. H. Cheng, G. J. Wang, and J. R. Sun, Chin. Phys. **7**, 741 (2002).

[9] A. Fujita, S. Fujieda, Y. Hasegawa, and K. Fukamichi, Phys. Rev. B **67**, 104416 (2003)

[10] H. Wada and Y. Tanabe, Appl. Phys. Lett. **79**, 3302 (2001).

[11] M. Balli, D. Fruchart, D. Gignoux, J. Tobola, E. K. Hlil, P. Wolfers, and R. Zach, J. Magn. Magn.Mater. **316**, 358 (2007).

[12] A. Plane, L. Mañosa, and M. Acet, J. Phys.: Condens. Matter **21**, 233201 (2009).

[13] S. Jacobs, J. Auringer, A. Boeder, J. Chell, L. Komorowski, J. Leonard, S. Russek, and C. Zimm, Int. J. Refrig. **37**, 84 (2014)

[14] A. Tura and A. Rowe, Int. J. Refrig. **34**, 628 (2011).

[15] M. Balli, P. Fournier, S. Jandl, M.M. Gospodinov, J. Appl. Phys. **115,** 173904(2014).

[16] M. Balli, B. Roberge, P. Fournier, S. Jandl, Crystals **7**, 44 (2017).

[17] C.R.H Bahl, D. Velasquz, K.K. Nielsen, K. Engelbrecht, K.B. Andersen, R. Bulatova, N. Pryds, Appl. Phys. Lett. **100**, 121905 (2012).

[18] D. Matte, M. de Lafontaine, A. Ouellet, M. Balli, P. Fournier, Phys. Rev. App. **9**, 054042 (2018).

[19] M. P. Singh, C. Grygiel, W. C. Sheets, Ph. Boullay, M. Hervieu, W. Prellier, B. Mercey, Ch. Simon and B. Raveau, Appl. Phys. Lett. **91**, 012503 (2007).

[20] C. Gauvin-Ndiaye, T. E. Baker, P. Karan, É. Massé, M. Balli, N. Brahiti, M. A. Eskandari, P. Fournier, A.-M. S. Tremblay and R. Nourafkan, Phys. Rev. B **98**, 125132 (2018).

[21] C. Gauvin-Ndiaye, A.-M. S. Tremblay, and R. Nourafkan, Phys. Rev B **99**, 125110 (2019).

[22] R. Shaheen, J. Bashir, M. Seddique, H. Rundlöf, A. Rennie, Physica B **385**,103 (2006).

[23] D. Kumar, V. Sudarshan, A.K. Singh, AIP Conf. Proc. **1953**, 080010 (2018).

[24] P. Barrozo, N.O.Moreno, J.A. Albino, Adv. Mater. Research **975**, 122 (2014).

[25] Sh.D. Bhame, V.L. J. Joly, P.A. Joy, Phys. Rev. B **72**, 054426 (2005).

[26] M. Balli, O.Sari, L.Zamni, C. Mahmed, J. Forchelet, Mater.Sci. Eng.B **177**, 629 (2012).

[27] M. Balli, P. Fournier, S. Jandl, K.D. Truong, M.M. Gospodinov, J.Appl. Phys. **116**, 073907 (2014).

[28] N.S. Rogada, J. Li, A.W. Sleight, M. A. Subramanian, Adv. Mater. **17**, 2225 (2005).

[29] X. Luo, Y. Sun, B.Wang, X. Zhu, W. Song, Z.,Yang, , J.Dai, Solid State Comm. **58**, 571 (2009).

[30] J. B. Goodenough, R. I. Dass, and J.-S. Zhou, Solid State Sci. **4**, 297 (2002).

[31] G. Blasse, J. Phys. Chem. Solids **26**, 1969 (1965).

[32] R. P. Madhogaria, R. Das, E. M. Clements, V. Kalappattil, M. H. Phan and H. Srikanth, Phys. Rev. B **99**, 104436 (2019).

[33] N. S. Bingham, M. H. Phan, H. Srikanth, M. A. Torija, and C. Leighton, J. Appl. Phys. **106**, 023909 (2009).

[34] D. Fatnassi, Kh. Sbissi, E. K. Hlil, M. Ellouze, J. L. Rehspringer, F. Elhalouani, Nanostruct. Chem. **5**, 375 (2015).

[35] M.E. Botello-Zubiate, M.C. Grijalva-Castillo, D. Soto-Parra, R. J. Sáenz-Hernández, C. R. Santillán-Rodríguez, and J. A. Matutes-Aquino, Materials **12**, 309 (2019).

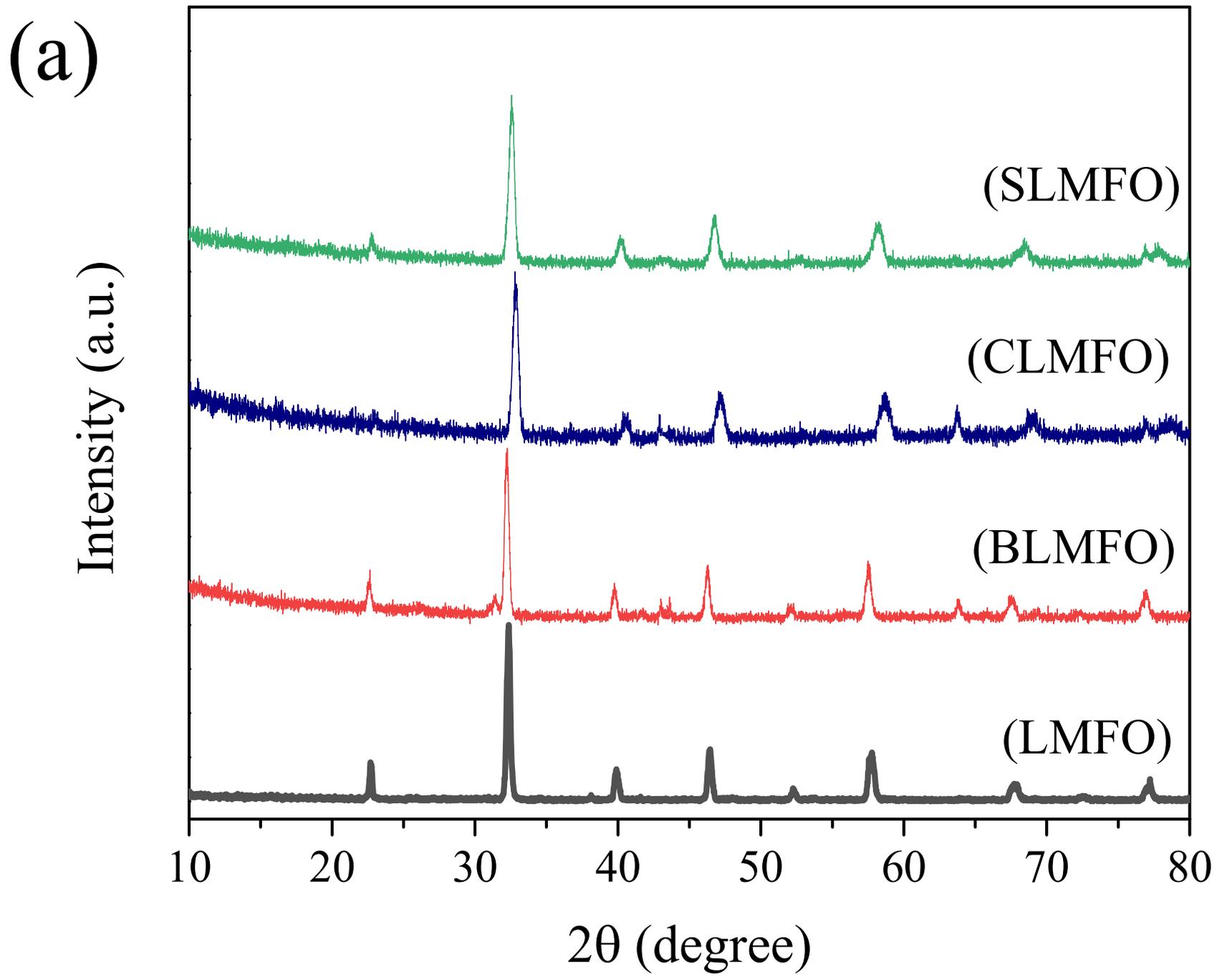

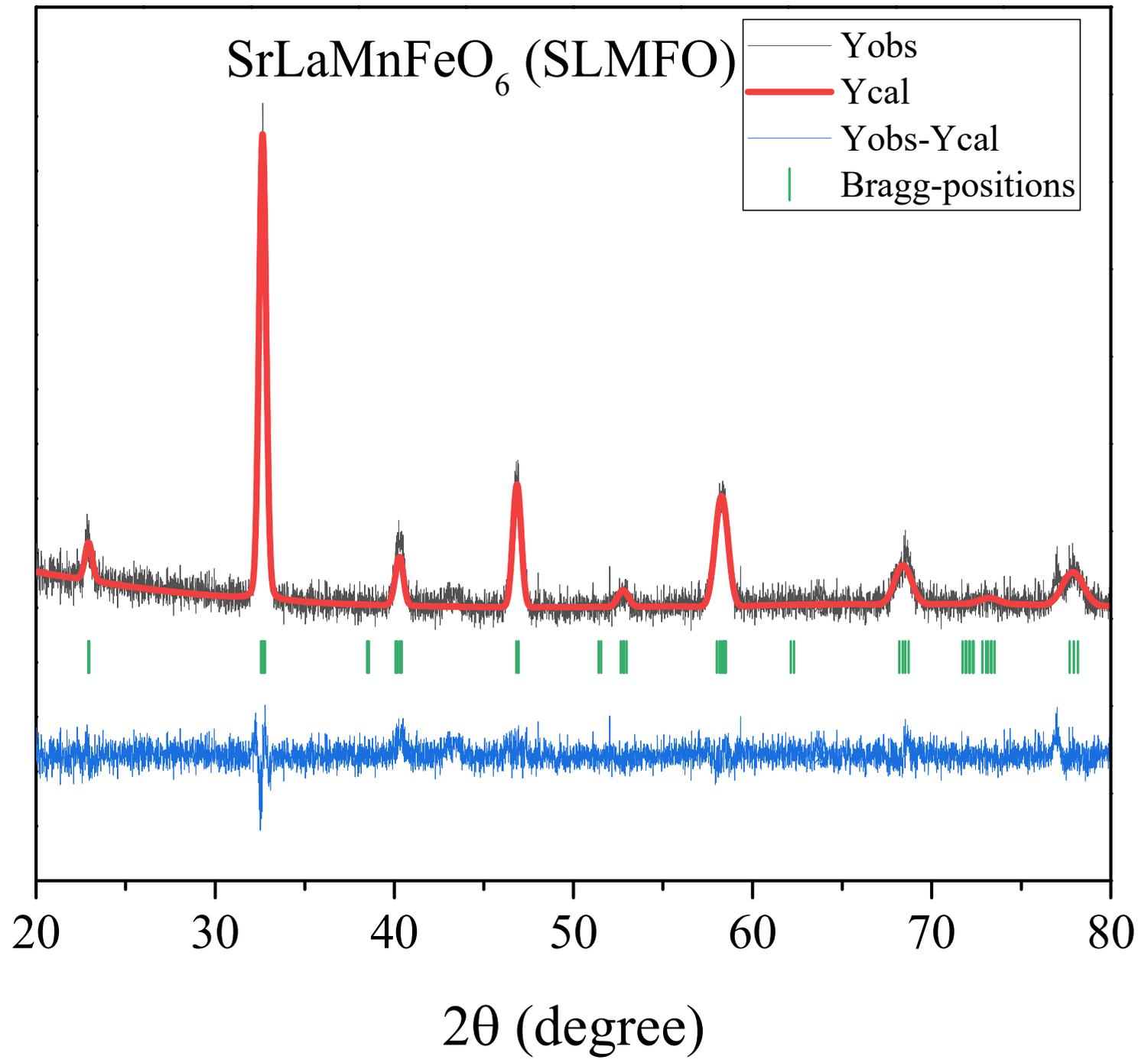

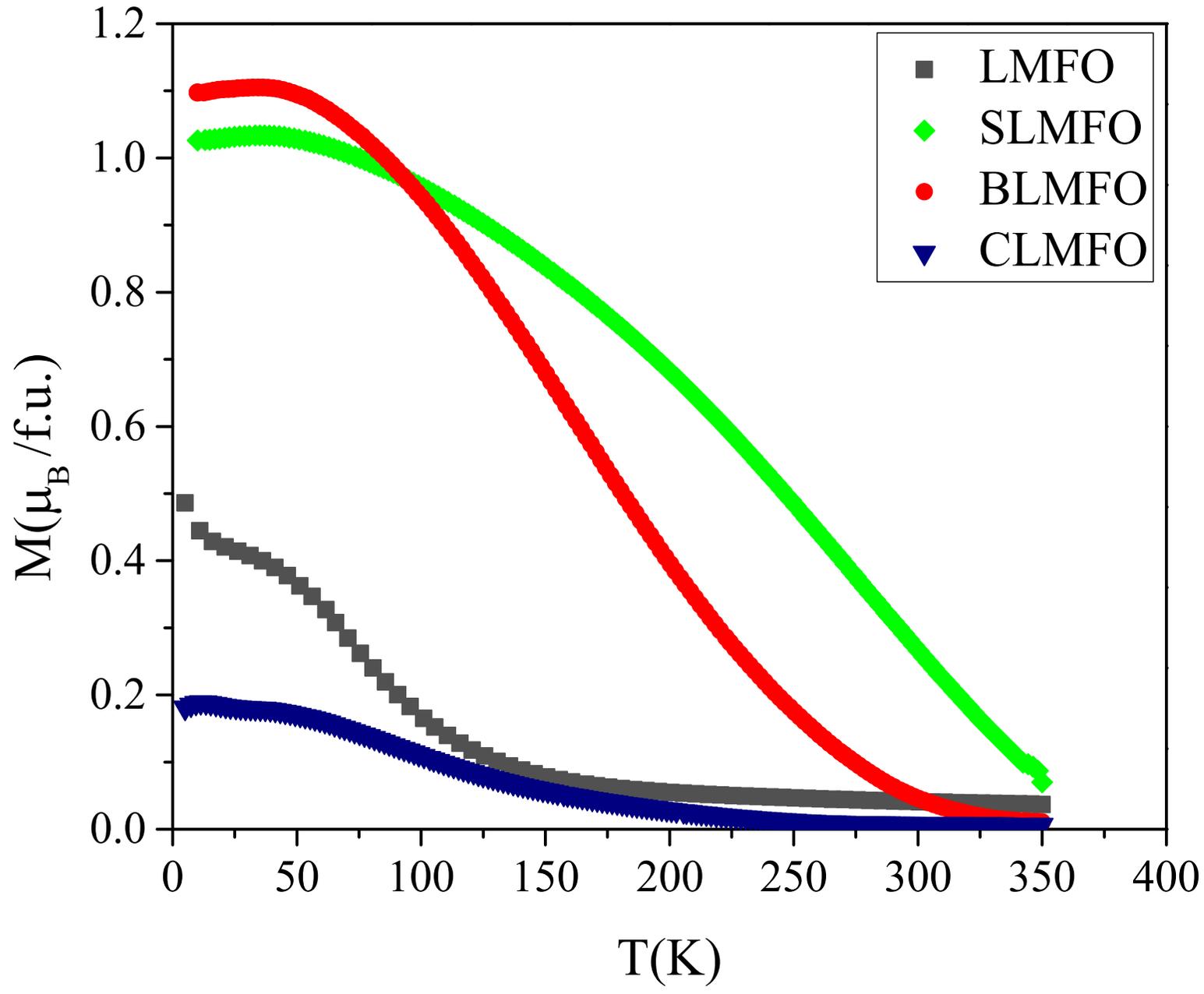

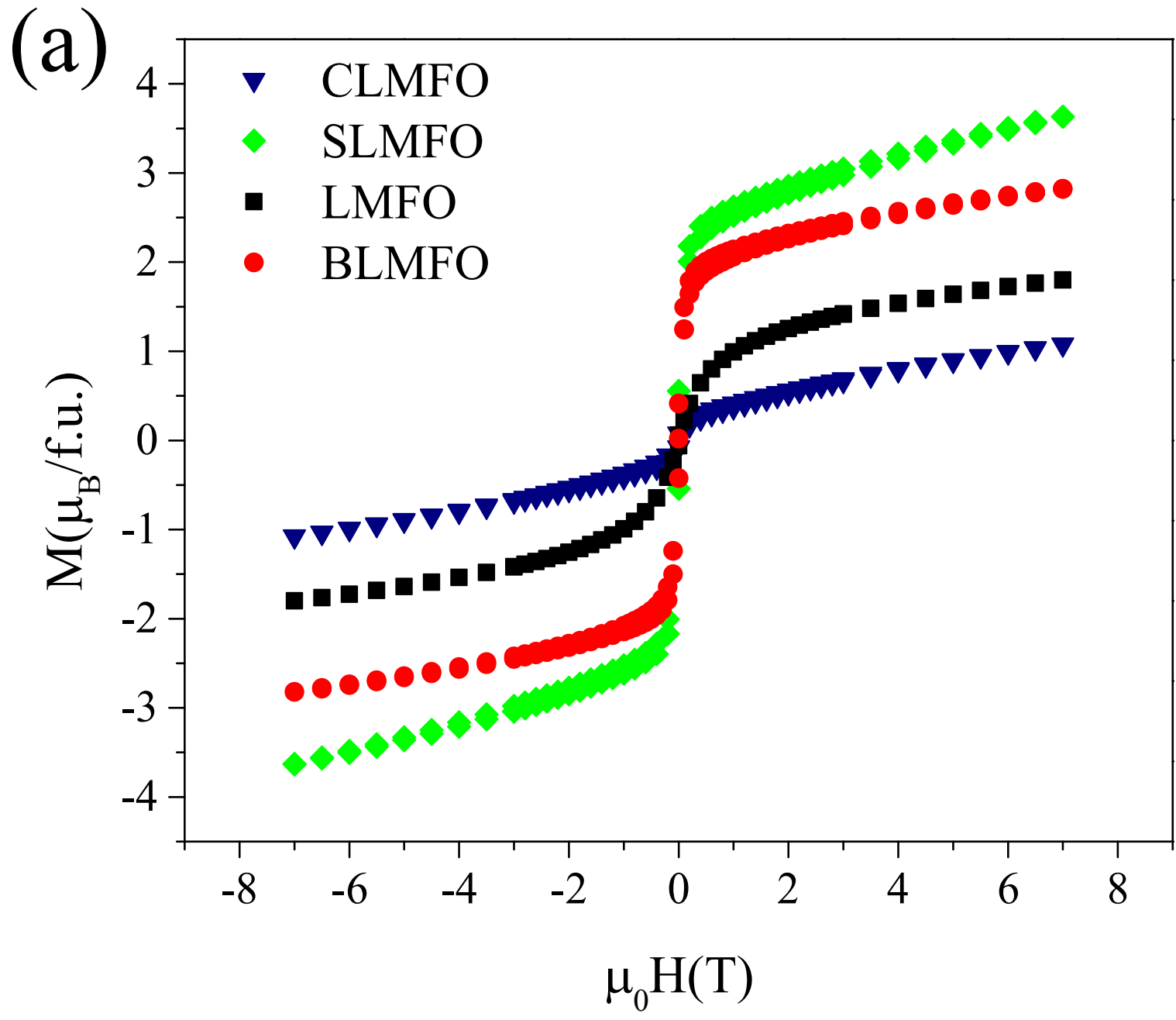

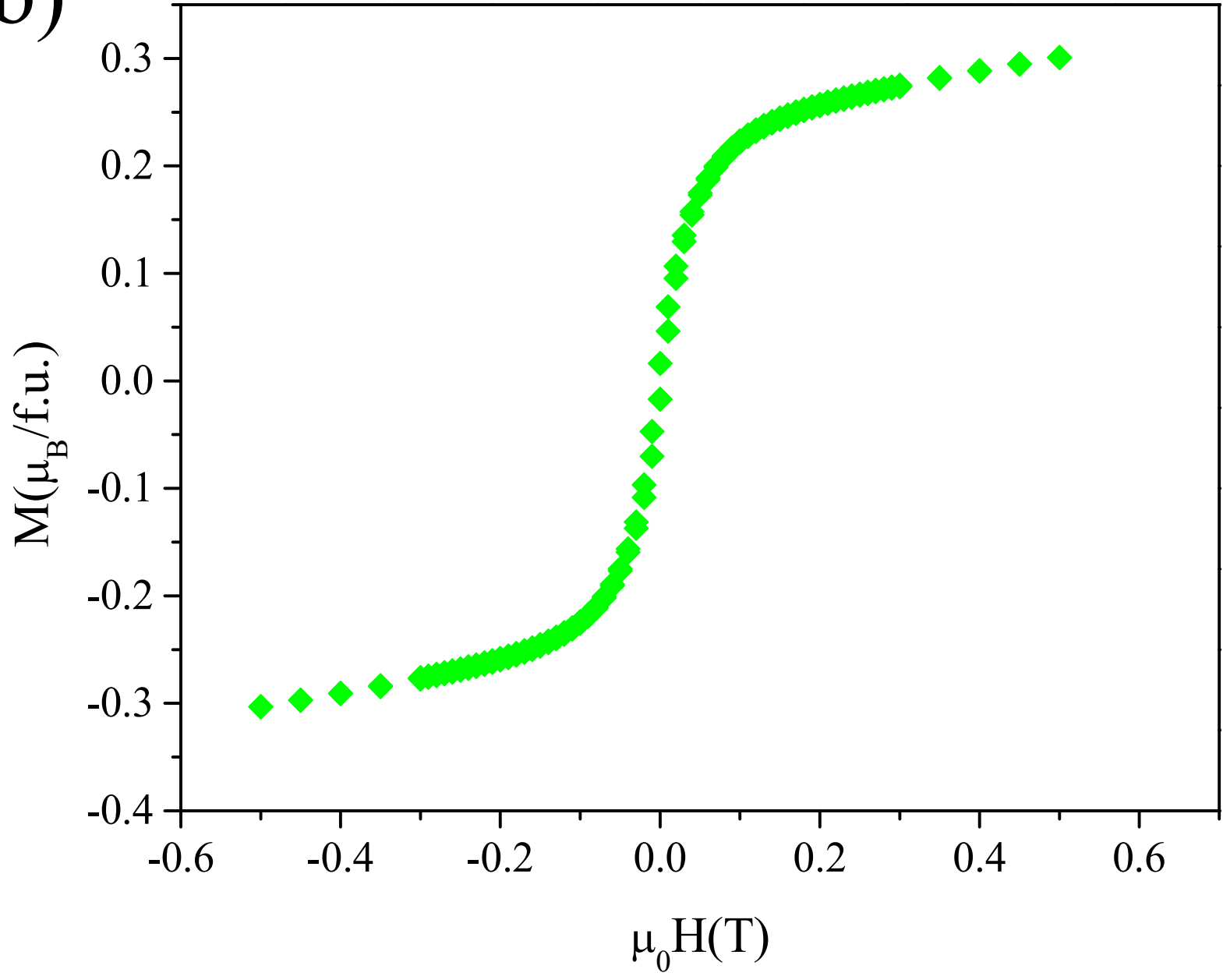

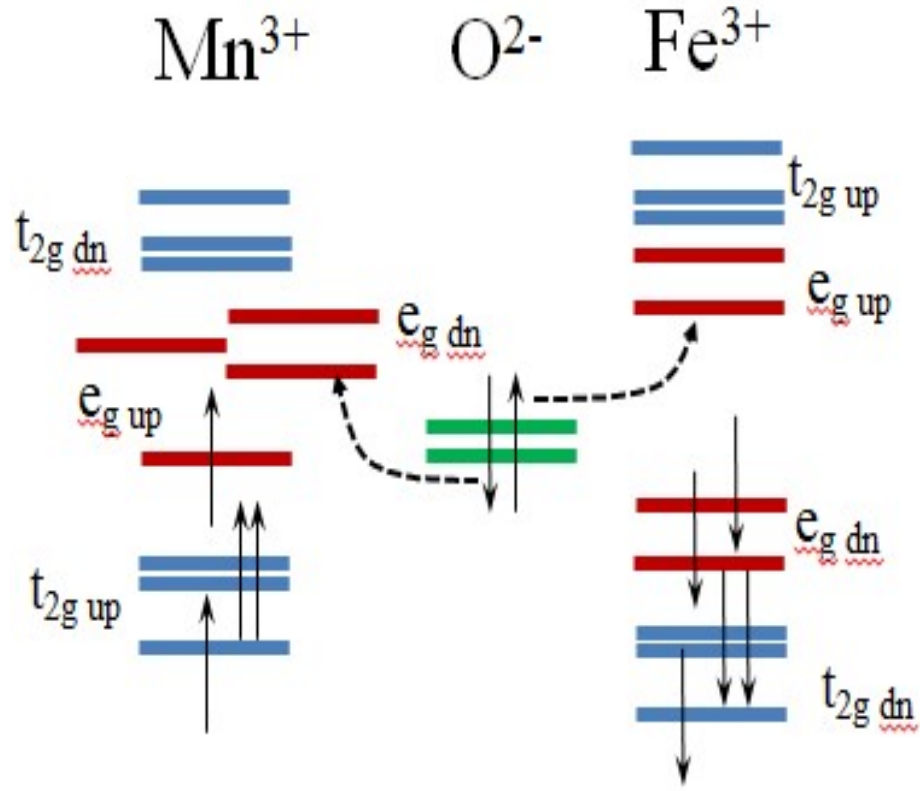
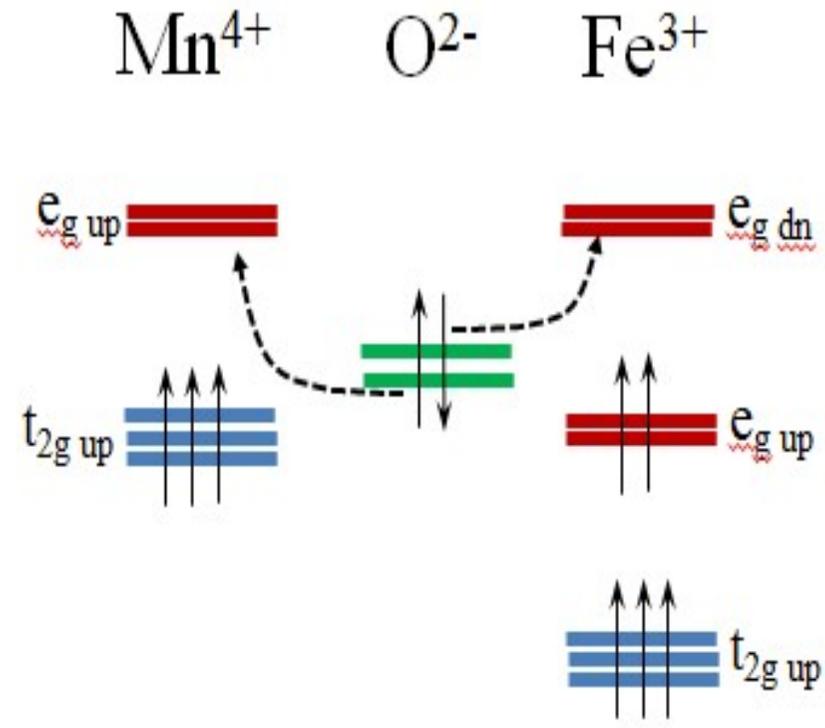

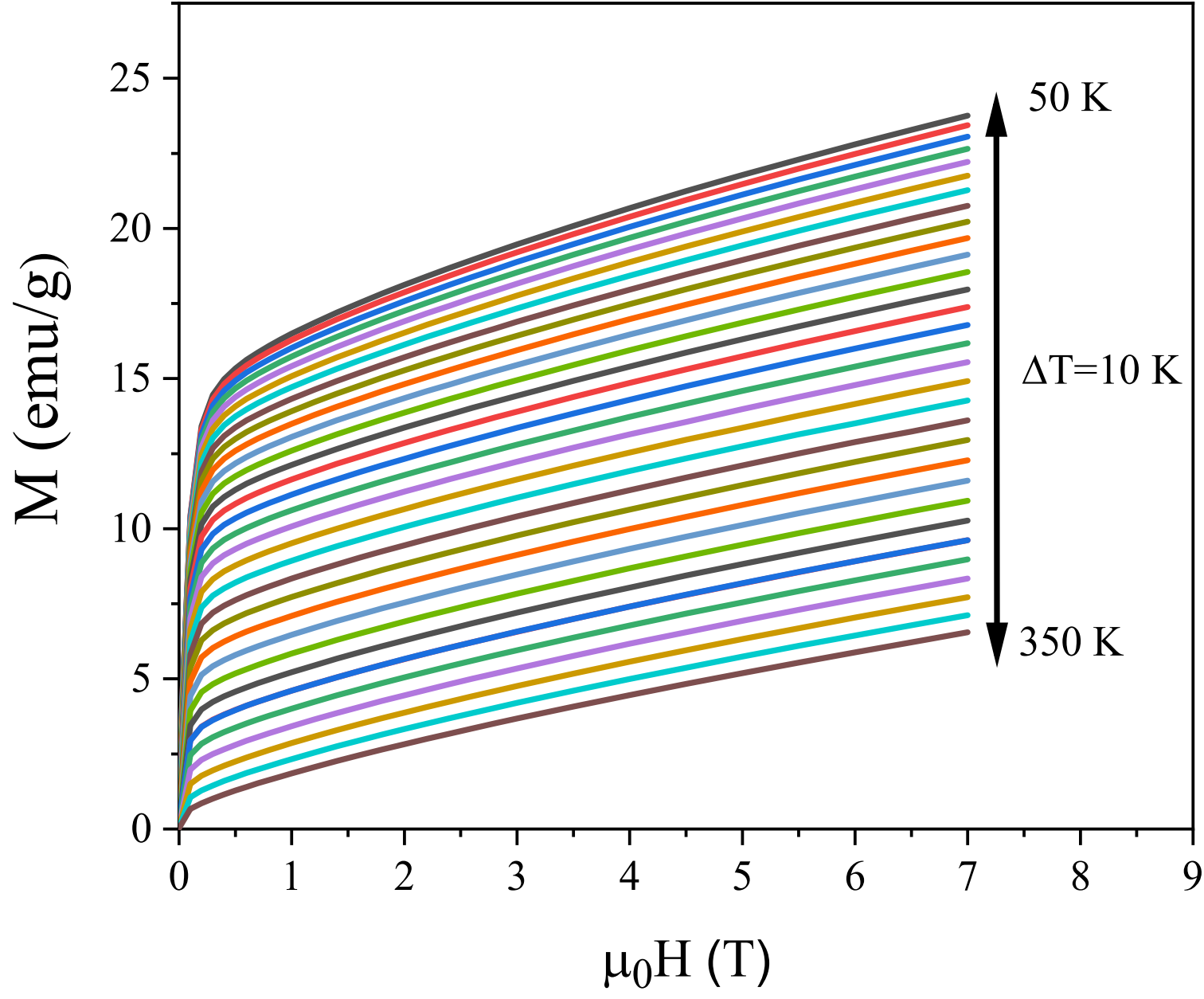

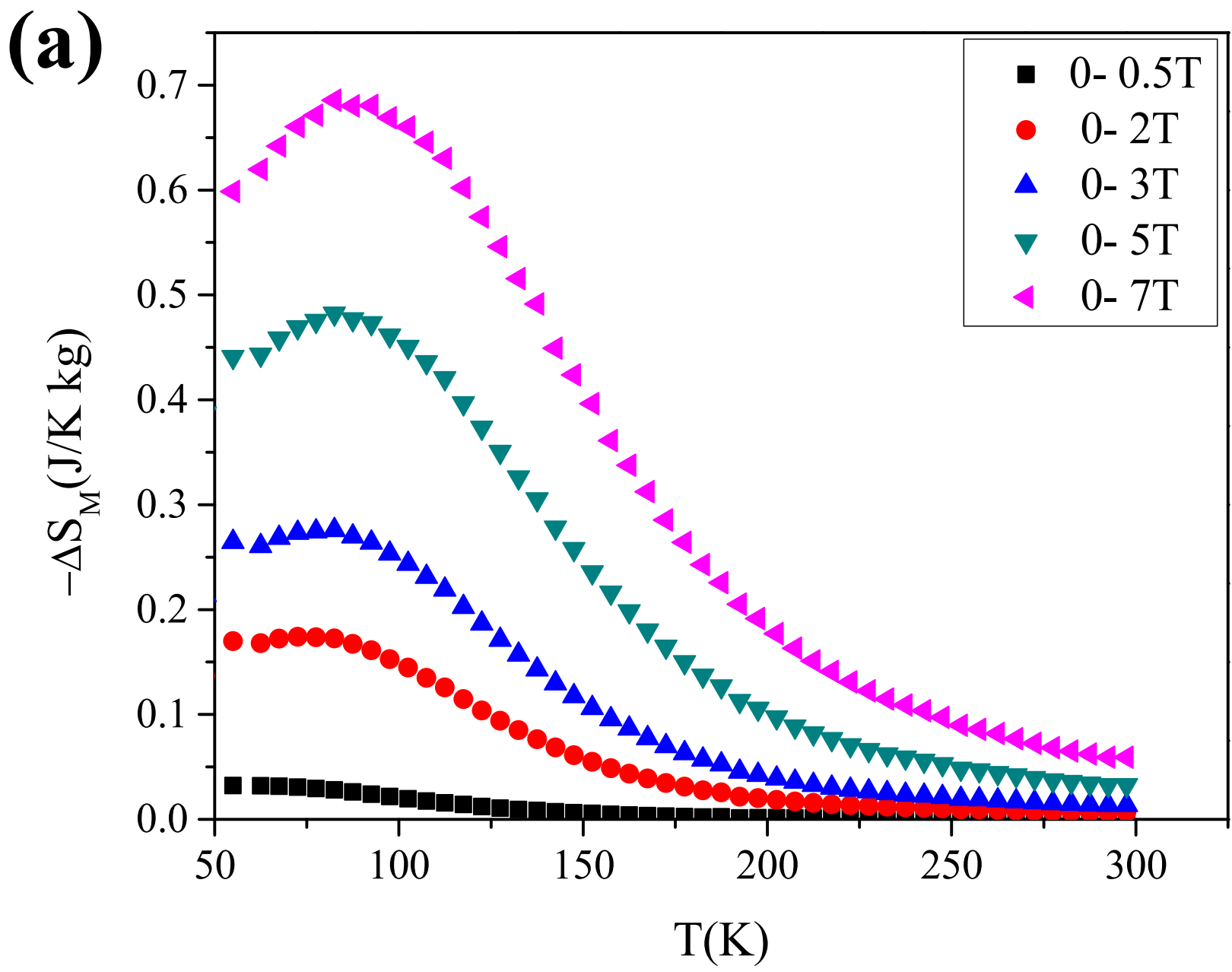

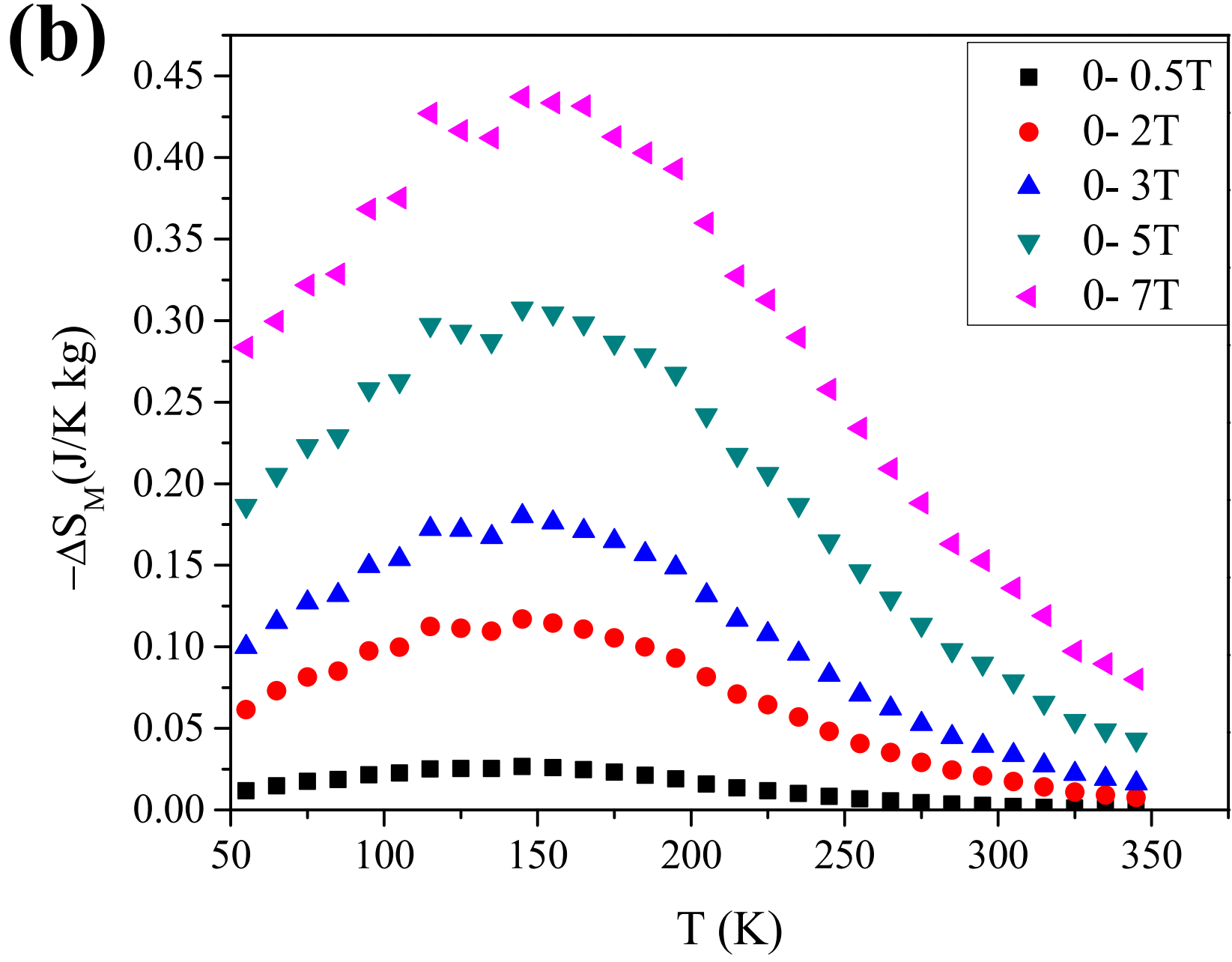

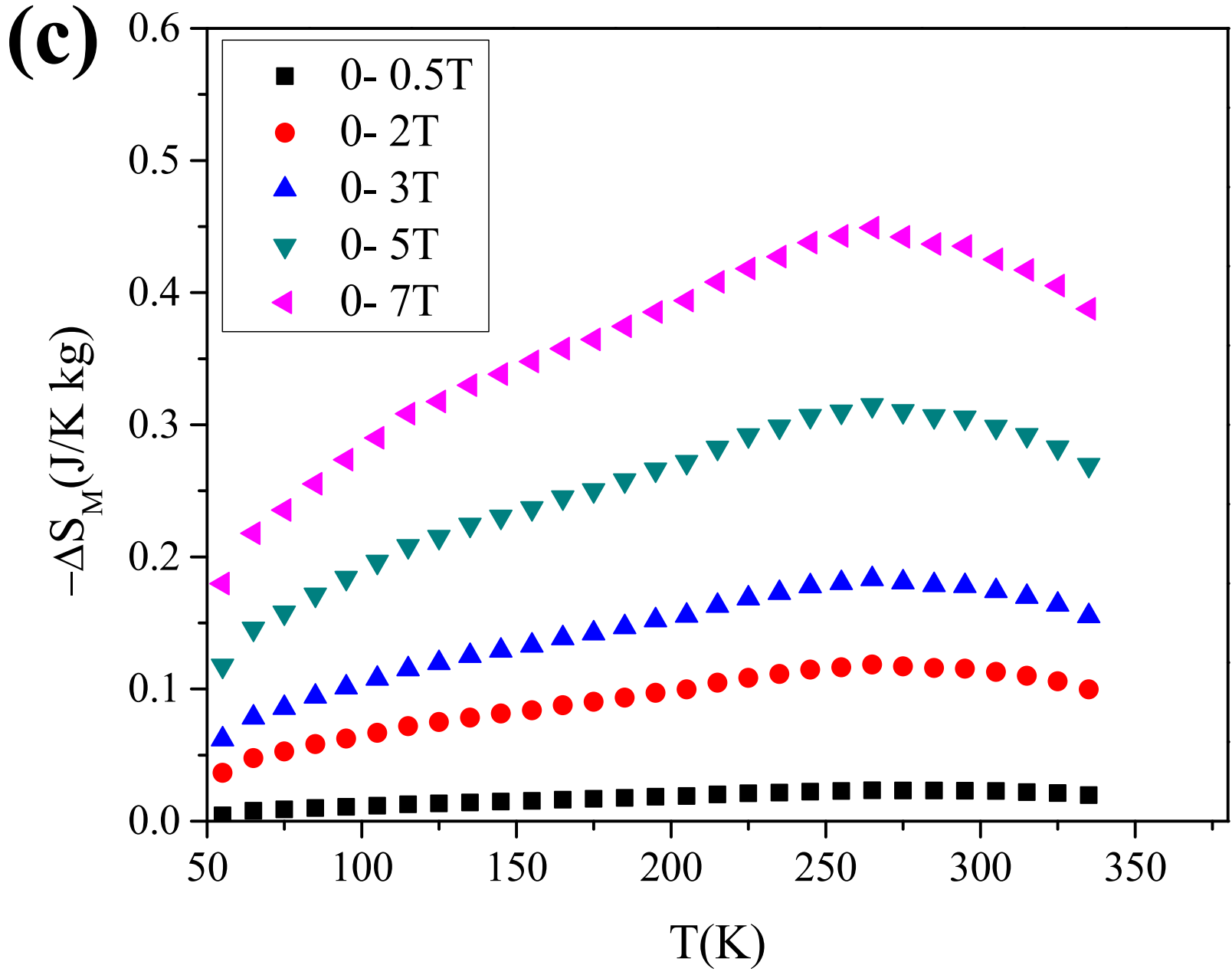

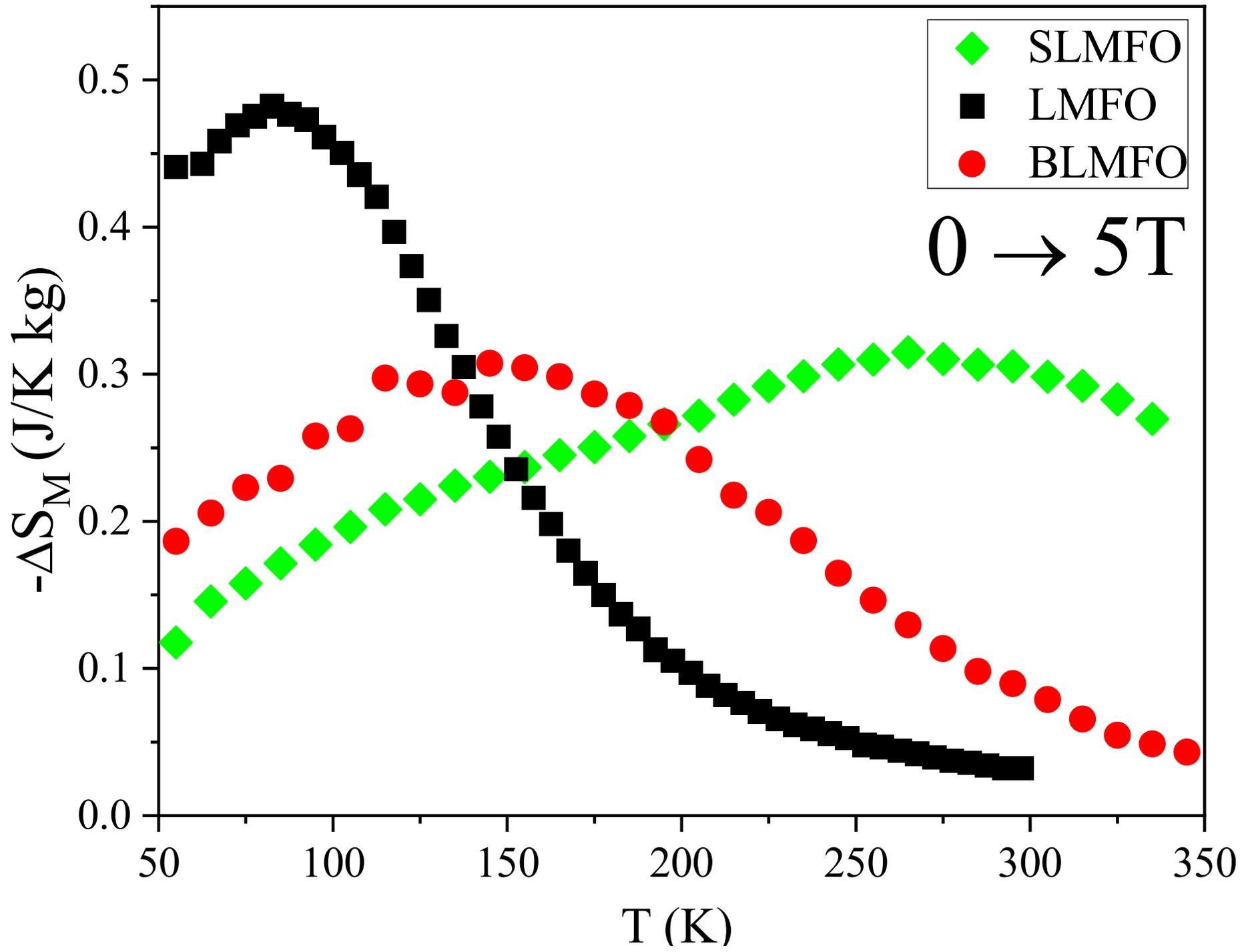

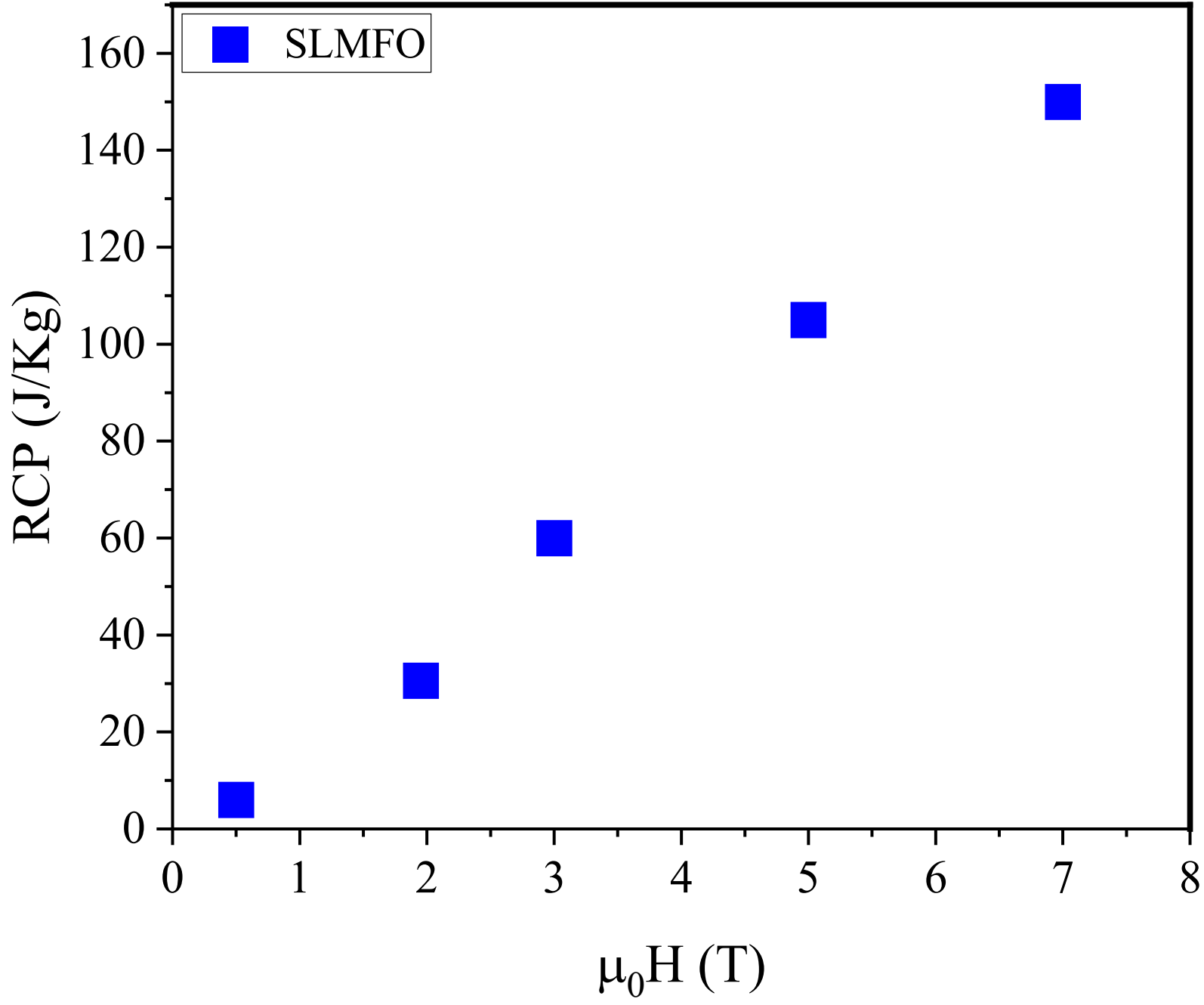

Supplementary material for :

# Analysis of the magnetic and magnetocaloric properties of $ALaFeMnO_6$ (A=Sr, Ba and Ca) double perovskites

N. Brahiti[1], M. Abbasi Eskandari[1], M. Balli[1,2], C. Gauvin-Ndiaye[1], R. Nourafkan[1], A.-M.S. Tremblay[1], P. Fournier[1,a]

[1]*Institut quantique, Regroupement québécois sur les matériaux de pointe et Département de physique, Université de Sherbrooke, Sherbrooke, J1K 2R1, Québec, Canada*

[2] *LERMA, ECINE, International University of Rabat, Parc Technopolis, Rocade de Rabat-Salé, 11100, Morocco*

a)Corresponding author and electronic mail: patrick.fournier@usherbrooke.ca

**Figure S-1** – Rietveld refinements for (a) LMFO, (b) BLMFO and (c) CLMFO. Black: experimental data; red: refinement fit; blue: difference; vertical lines: markers of predicted reflections.

The structural parameters were obtained by fitting the experimental XRD data using the Rietveld structural refinement FULLPROF software applying the pseudo – Voigt peak shape function and a linear interpolation for background description. The refinements were performed until convergence as shown by the goodness of fit ($\chi^2$).

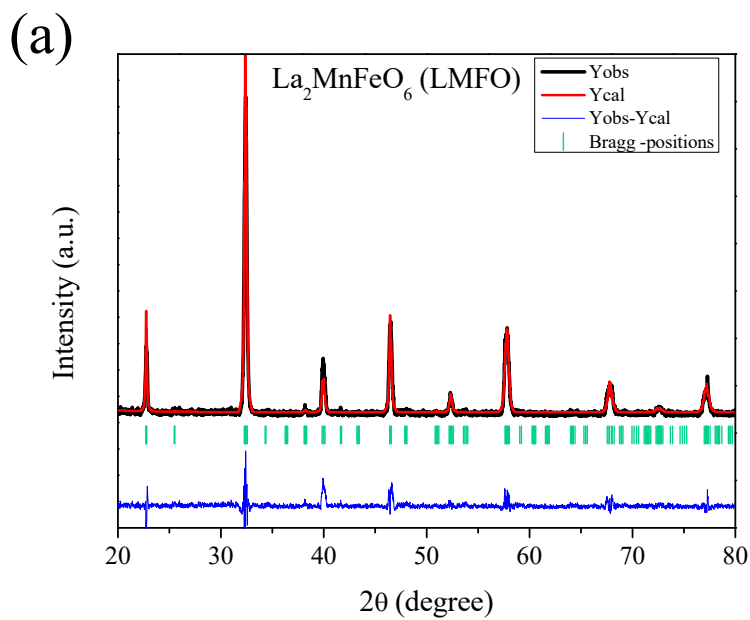

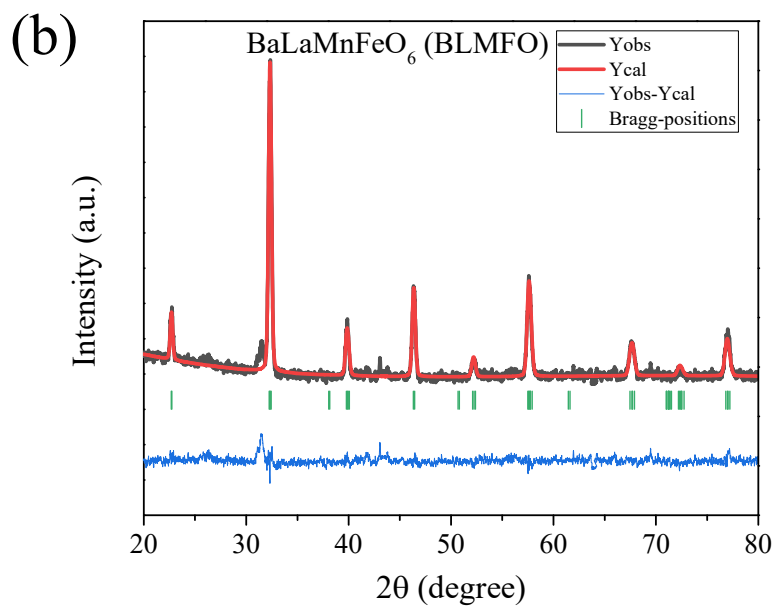

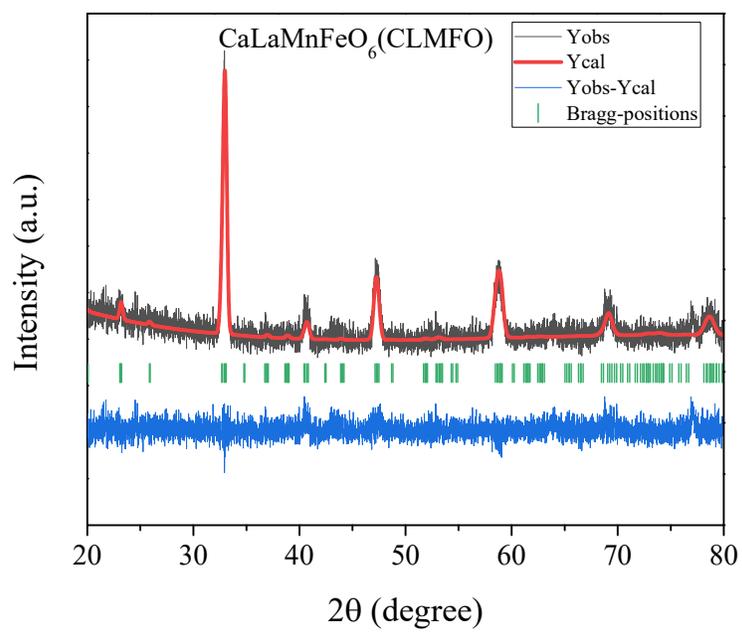

**Figure S-2** -Isothermal magnetization curves for (a) BLMFO and (b) LMFO samples from 10 K to 350 K used to evaluate the isothermal entropy change in Figure 6. Isotherms were measured at a fixed temperature interval of 10K. The data in (a) were measured in a Physical properties measurements system from Quantum Design using the ACMS option up to 9T.

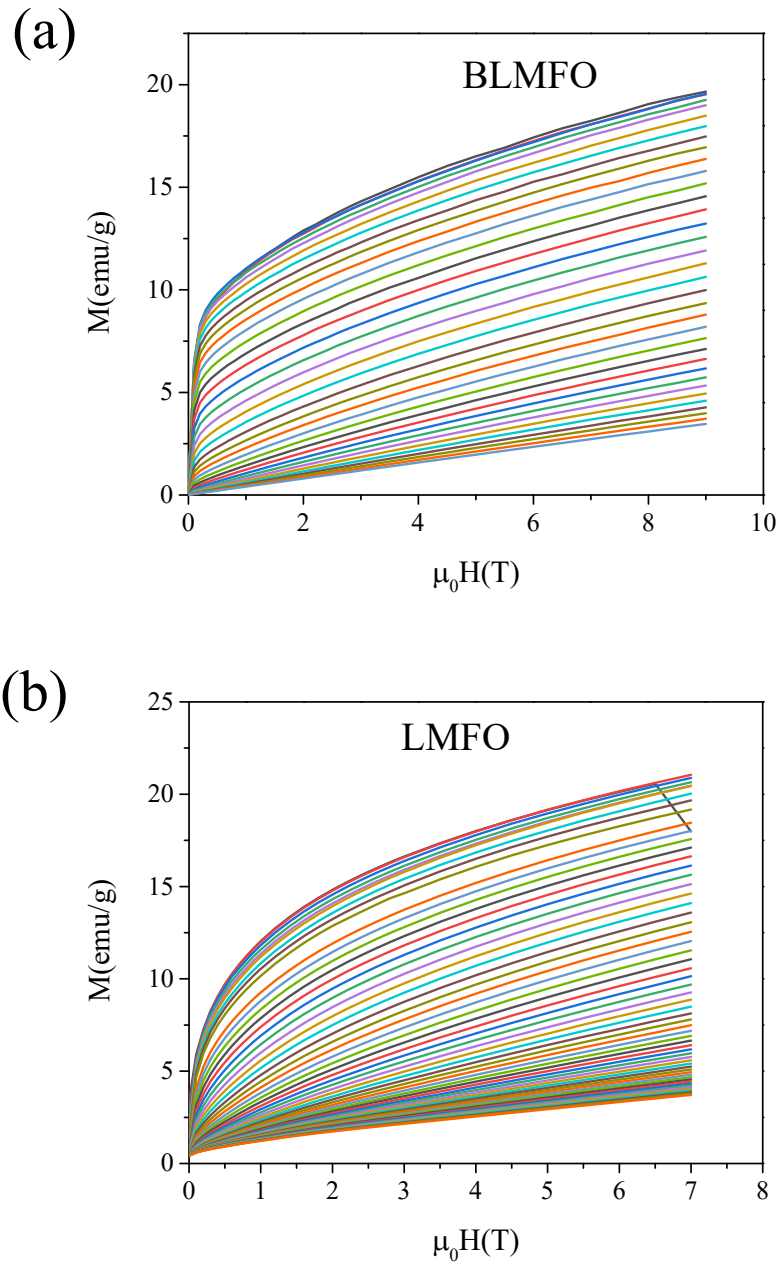